\documentstyle[12pt]{article}

\begin{document}

\author{C. Bizdadea\thanks{%
e-mail address: bizdadea@hotmail.com}\ and S. O. Saliu\thanks{%
e-mail address: osaliu@central.ucv.ro} \\
Department of Physics, University of Craiova\\
13 A.I. Cuza Str., Craiova RO-1100, Romania}
\title{Irreducible antifield formalism for reducible constrained
Hamiltonian systems}
\maketitle

\begin{abstract}
Reducible constrained Hamiltonian systems are quantized accordingly an
irreducible BRST manner. Our procedure is based on the construction of an
irreducible theory which is physically equivalent with the original one. The
equivalence between the two systems makes legitimate the substitution of the
BRST quantization for the reducible theory by that of the irreducible
system. The general formalism is illustrated in detail on a model involving
abelian one- and two-form gauge fields.

\end{abstract}

\section{Introduction}

It is well-known that there are two BRST approaches to the quantization of
arbitrary gauge theories. One of them is based on the Lagrangian formalism 
\cite{1}--\cite{6} (known as the antifield formalism), while the other is
dealing with Hamiltonian aspects \cite{6}--\cite{11}. Both formulations can
be applied to irreducible, as well as reducible gauge systems. For reducible
theories, it is necessary to introduce ghosts of ghosts and their antifields
in order to ensure the nilpotency of the BRST symmetry. The antifield
treatment was extended to constrained Hamiltonian systems \cite{12}--\cite
{13}, allowing therefore a clearer connection between the Lagrangian and the
Hamiltonian BRST symmetries.

In this paper we give a consistent procedure for quantizing reducible
Hamiltonian systems with first-class constraints following an irreducible
BRST mechanism. Although the idea of replacing a redundant set of
first-class constraints by an irreducible one in a larger phase-space is
known \cite{6}, \cite{14}, it has been neither consistently developed nor
applied so far to the quantization of reducible gauge theories. Starting
with a finite-stage reducible Hamiltonian first-class system, we perform the
following steps: (i) we transform the original reducible theory into an
irreducible one in a manner that allows the substitution of the BRST
quantization of the reducible system by that of the irreducible theory, and
(ii) we quantize the extended action of the irreducible system accordingly
the antifield-BRST formalism. In consequence, the ghosts of ghosts, as well
as their antifields do not appear within our formalism. By virtue of this,
our method puts on equal footing the reducible and irreducible constrained
Hamiltonian systems from the BRST formalism point of view. As far as we
know, such an approach has not been previously published, hence our paper
establishes a new result.

The paper is organized in six sections. Section 2 is dealing with enlarging
the initial phase-space of an arbitrary first-stage reducible Hamiltonian
system by adding some supplementary canonical pairs, and with further
constructing an irreducible set of first-class constraints. The irreducible
set is derived in a way that ensures the equivalence with the starting
first-class set. In Section 3 we establish the physical equivalence between
the reducible and irreducible systems. In this light, the physical
observables and also the number of physical degrees of freedom associated
with both theories are shown to coincide. The physical equivalence allows
the replacement of the reducible BRST quantization by the irreducible one.
The quantization of the resulting irreducible first-class Hamiltonian system
is then performed on behalf of an appropriate gauge-fixing fermion. Section
4 exposes the generalization of our results from the previous sections to
finite-stage reducible first-class constraints. In Section 5 we exemplify
the general theory in the case of a reducible model describing the
Stueckelberg coupling between abelian one- and two-form gauge fields.
Section 6 ends the paper with some conclusions.

\section{First-stage reducible Hamiltonian theories}

In this section we show how one can construct a set of irreducible
first-class constraints starting from a first-stage reducible one. We begin
with a system described by $N$ canonical pairs $\left( q^i,p_i\right) $,
subject to the first-class constraints 
\begin{equation}
\label{1}G_{a_0}\left( q,p\right) \approx 0,\;a_0=1,\ldots ,M_0, 
\end{equation}
which are assumed to be first-stage reducible 
\begin{equation}
\label{2}Z_{\;\;a_1}^{a_0}G_{a_0}=0,\;a_1=1,\ldots ,M_1, 
\end{equation}
and suppose that there are no second-class constraints in the theory. In (%
\ref{2}) we used the strong equality because one can always define the
first-stage reducibility functions such that to have off-shell reducibility.
For the sake of simplicity, we assume that $\left( q^i,p_i\right) $ are
bosonic, but the results can be extended to fermions by introducing some
appropriate phases. We denote the first-class Hamiltonian by $H$, such that
the gauge algebra is expressed by $\left[ G_{a_0},G_{b_0}\right]
=C_{\;\;a_0b_0}^{c_0}G_{c_0}$, $\left[ H,G_{a_0}\right]
=V_{\;\;a_0}^{b_0}G_{b_0}$. Relations (\ref{2}) indicate that the functions $%
G_{a_0}$ are not all independent. Under these circumstances, we locally
split these functions within the independent and dependent components, $%
G_{\bar a_0}$, respectively, $G_{a_1}$ 
\begin{equation}
\label{3}G_{a_0}=\left( 
\begin{array}{c}
G_{\bar a_0} \\ 
G_{a_1} 
\end{array}
\right) ,\bar a_0=1,\ldots ,M_0-M_1, 
\end{equation}
with 
\begin{equation}
\label{4}G_{a_1}=M_{a_1}^{\;\;\bar a_0}G_{\bar a_0}, 
\end{equation}
for some functions $M_{a_1}^{\;\;\bar a_0}$, such that $G_{\bar a_0}\approx
0\Rightarrow G_{a_1}\approx 0$. All that is required is to choose the
functions $G_{a_0}$ in such a way that the split can be achieved in
principle. With the help of (\ref{4}) we solve (\ref{2}) with respect to $%
Z_{\;\;a_1}^{a_0}$.Accordingly, we find 
\begin{equation}
\label{5}Z_{\;\;b_1}^{a_0}=\left( M_{b_1}^{\;\;\bar a_0},-\delta
_{b_1}^{\;\;a_1}\right) . 
\end{equation}
Next, we perform a transformation 
\begin{equation}
\label{6}G_{a_0}\rightarrow \tilde G_{a_0}=\left( 
\begin{array}{c}
G_{\bar a_0} \\ 
{\bf 0} 
\end{array}
\right) , 
\end{equation}
with the help of an invertible matrix $M_{a_0}^{\;\;b_0}$, 
\begin{equation}
\label{7}\tilde G_{a_0}=M_{a_0}^{\;\;b_0}G_{b_0}, 
\end{equation}
such that $\tilde G_{a_0}\approx 0\Leftrightarrow G_{a_0}\approx 0$. This
matrix allows the representation 
\begin{equation}
\label{8}M_{a_0}^{\;\;b_0}=\left( 
\begin{array}{cc}
\delta _{\bar a_0}^{\;\;\bar b_0} & {\bf 0} \\ M_{a_1}^{\;\;\bar b_0} & 
-\delta _{a_1}^{\;\;b_1} 
\end{array}
\right) , 
\end{equation}
while its inverse coincides with itself 
\begin{equation}
\label{9}\bar M_{a_0}^{\;\;b_0}=M_{a_0}^{\;\;b_0}. 
\end{equation}
If one inverses (\ref{7}), one gets 
\begin{equation}
\label{10}G_{a_0}=M_{a_0}^{\;\;b_0}\tilde G_{b_0}, 
\end{equation}
so, on account of (\ref{10}) and (\ref{2}), we consequently find 
\begin{equation}
\label{11}Z_{\;\;b_1}^{a_0}G_{a_0}=Z_{\;\;b_1}^{a_0}M_{a_0}^{\;\;b_0}\tilde
G_{b_0}=0. 
\end{equation}
In this way, we can regard 
\begin{equation}
\label{12}\tilde Z_{\;\;c_1}^{b_0}=Z_{\;\;c_1}^{a_0}M_{a_0}^{\;\;b_0}, 
\end{equation}
as the reducibility functions of $\tilde G_{b_0}$. Using (\ref{5}) and (\ref
{8}) it follows that $\tilde Z_{\;\;c_1}^{b_0}$ is given by 
\begin{equation}
\label{13}\tilde Z_{\;\;c_1}^{b_0}=\left( {\bf 0},\delta
_{c_1}^{\;\;b_1}\right) . 
\end{equation}
If one splits the free index in (\ref{12}) like $b_0=\left( \bar
b_0,b_1\right) $ and uses (\ref{13}), one derives (for $b_0\rightarrow b_1$) 
\begin{equation}
\label{14}Z_{\;\;c_1}^{a_0}M_{a_0}^{\;\;b_1}=\delta _{c_1}^{\;\;b_1}, 
\end{equation}
hence 
\begin{equation}
\label{15}rank\left( Z_{\;\;c_1}^{a_0}M_{a_0}^{\;\;b_1}\right) =M_1, 
\end{equation}
where 
\begin{equation}
\label{16}M_{a_0}^{\;\;b_1}=\left( 
\begin{array}{c}
0 \\ 
-\delta _{a_1}^{\;\;b_1} 
\end{array}
\right) . 
\end{equation}

Next, we transform the reducible constraints (\ref{1}) into some irreducible
ones. In this respect, we introduce a canonical pair $\left( y^{a_1},\pi
_{a_1}\right) $ associated with every (free index of) relation (\ref{2}),
which we impose to be constrained by 
\begin{equation}
\label{17}\pi _{a_1}\approx 0. 
\end{equation}
Obviously, the constraints (\ref{1}) and (\ref{17}) are first-class and
reducible. The theory based on these constraints is physically equivalent
with that based only on the constraints (\ref{1}) as the two systems display
the same number of physical degrees of freedom, and, moreover, it can be
shown that they describe the same physical observables. Indeed, if $f$
denotes an observable of the theory with the constraints (\ref{1}) and (\ref
{17}), then it is also an observable of the original one. The last statement
arises in a simple manner by writing down the equations satisfied by $f$,
namely, 
\begin{equation}
\label{17a}\left[ f,G_{a_0}\right] \approx 0,\;\left[ f,\pi _{a_1}\right]
\approx 0. 
\end{equation}
The equations (\ref{17a}) show that $f$ does not depend (at least weakly) on 
$y^{a_1}$, and, in addition, that the observables associated with this
theory fulfill $\left[ f,G_{a_0}\right] \approx 0$, which are nothing but
the equations verified by the observables corresponding to the original
system. The converse is also valid, i.e., any observable of the original
theory satisfies (\ref{17a}) because it does not depend on $\left(
y^{a_1},\pi _{a_1}\right) $ and checks by definition $\left[
f,G_{a_0}\right] \approx 0$.

Dropping out the trivial part of (\ref{6}), we construct the irreducible
first-class constraints 
\begin{equation}
\label{18}\tilde \gamma _{a_0}=\left( 
\begin{array}{c}
G_{\bar a_0} \\ 
\pi _{a_1} 
\end{array}
\right) \approx 0, 
\end{equation}
such that the momenta $\pi _{a_1}$ replace the dependent constraint
functions. With the help of (\ref{6}) and (\ref{16}), the constraints (\ref
{18}) can be put under the form 
\begin{equation}
\label{19}\tilde \gamma _{a_0}=\tilde G_{a_0}-M_{a_0}^{\;\;b_1}\pi
_{b_1}\approx 0. 
\end{equation}
Now, we pass from (\ref{19}) to the equivalent set of first-class
constraints 
\begin{equation}
\label{20}\gamma _{a_0}=M_{a_0}^{\;\;b_0}\tilde \gamma _{b_0}, 
\end{equation}
with $M_{a_0}^{\;\;b_0}$ the matrix (\ref{8}). Making the notations $%
A_{a_0}^{\;\;\,b_1}=-M_{a_0}^{\;\;b_0}M_{b_0}^{\;\;b_1}$ and using (\ref{10}%
), we find from (\ref{20}) the first-class constraints 
\begin{equation}
\label{21}\gamma _{a_0}\equiv G_{a_0}+A_{a_0}^{\;\;b_1}\pi _{b_1}\approx 0. 
\end{equation}
The matrix $A_{a_0}^{\;\;b_1}$ also verifies (\ref{14}). Indeed, we have
that 
\begin{equation}
\label{22}Z_{\;\;c_1}^{a_0}A_{a_0}^{\;\;b_1}=-Z_{\;\;c_1}^{a_0}M_{a_0}^{\;%
\;b_0}M_{b_0}^{\;\;b_1}=\tilde Z_{\;\;c_1}^{b_0}M_{b_0}^{\;\;b_1}=\delta
_{c_1}^{\;\;b_1}. 
\end{equation}
However, from practical reasons it is useful to weaken the condition (\ref
{22}) by taking $A_{a_0}^{\;\;b_1}$ such that $Z_{\;\;c_1}^{a_0}A_{a_0}^{\;%
\;b_1}=D_{c_1}^{\;\;b_1}$ is invertible, i.e., 
\begin{equation}
\label{23}rank\left( D_{c_1}^{\;\;b_1}\right) =M_1. 
\end{equation}
We employ this choice throughout the paper. Moreover, the first-class
constraints (\ref{21}) are irreducible. Indeed, we have that $%
Z_{\;\;a_1}^{a_0}\gamma _{a_0}=D_{a_1}^{\;\;b_1}\pi _{b_1}$ is non-vanishing
due to (\ref{23}).

Within the above discussion we supposed that the split of the reducible
constraints into independent and dependent ones can be done in principle,
this assumption being useful for some technical purposes. Indeed, the split
form of the original constraints represents an intermediate step in finally
reaching the irreducible constraints (\ref{21}) where the initial constraint
functions appear in a covariant (not split) form. The derivation of the
constraints (\ref{21}) based on the above split is still useful in order to
evidence the introduction of the intermediate reducible system possessing
the constraints (\ref{1}) and (\ref{17}), which subsequently emphasizes in a
suggestive manner how the dependent constraints can be replaced by some new
degrees of freedom ensuring the irreducibility. However, the separation of
the reducible constraints can spoil the covariance or destroy the locality
of those relations where it is manifest. In fact, the split hypothesis is
not crucial in arriving at (\ref{21}) and can be replaced by homological
arguments, as follows. It is well-known that the BRST symmetry $s_R$
associated with a Hamiltonian reducible theory contains two basic
differentials 
\begin{equation}
\label{23a}s_R=\delta _R+D_R+\cdots ,
\end{equation}
where $\delta _R$ denotes the Koszul-Tate differential and $D_R$ stands for
a model of longitudinal derivative along the gauge orbits. In the case of
first-stage reducible systems, the action of $\delta _R$ on the original
phase-space variables and on the generators $\left( {\cal P}_{a_0},{\cal P}%
_{a_1}\right) $ in the Koszul-Tate complex reads as 
\begin{equation}
\label{23b}\delta _Rq^i=0,\;\delta _Rp_i=0,
\end{equation}
\begin{equation}
\label{23c}\delta _R{\cal P}_{a_0}=-G_{a_0},
\end{equation}
\begin{equation}
\label{23d}\delta _R{\cal P}_{a_1}=-Z_{\;\;a_1}^{a_0}{\cal P}_{a_0},
\end{equation}
with ${\cal P}_{a_0}$ and ${\cal P}_{a_1}$ of antighost number one,
respectively, two. The antighosts ${\cal P}_{a_1}$ are required in order to
kill the antighost number one non-trivial co-cycles 
\begin{equation}
\label{23e}\rho _{a_1}\equiv Z_{\;\;a_1}^{a_0}{\cal P}_{a_0},
\end{equation}
in the homology of $\delta _R$. The idea with the help of which we can
recover (\ref{21}) is to redefine the antighosts ${\cal P}_{a_0}$ such that
the non-trivial co-cycles of the type (\ref{23e}) vanish identically. If we
succeed in doing this, the co-cycles (\ref{23e}) do not appear anymore,
hence the antighosts ${\cal P}_{a_1}$ are no longer necessary such that the
theory becomes indeed irreducible. In this light, we perform the
transformation 
\begin{equation}
\label{23f}{\cal P}_{a_0}\rightarrow \stackrel{\sim }{\cal P}%
_{a_0}=D_{a_0}^{\;\;b_0}{\cal P}_{b_0},
\end{equation}
where $D_{a_0}^{\;\;b_0}$ is chosen such that 
\begin{equation}
\label{23g}Z_{\;\;a_1}^{a_0}D_{a_0}^{\;\;b_0}=0,\;D_{a_0}^{\;%
\;b_0}G_{b_0}=G_{a_0}.
\end{equation}
From (\ref{23c}) and (\ref{23f}--\ref{23g}) we obtain that 
\begin{equation}
\label{23h}\delta \stackrel{\sim }{\cal P}_{a_0}=-G_{a_0},
\end{equation}
which subsequently leads to 
\begin{equation}
\label{23i}\delta \left( Z_{\;\;a_1}^{a_0}\stackrel{\sim }{\cal P}%
_{a_0}\right) =0,
\end{equation}
but with 
\begin{equation}
\label{23j}Z_{\;\;a_1}^{a_0}\stackrel{\sim }{\cal P}_{a_0}\equiv 0.
\end{equation}
In (\ref{23h}--\ref{23i}) we redenoted $\delta _R$ by $\delta $ in order to
outline that the new theory is irreducible. If we take 
\begin{equation}
\label{23k}D_{a_0}^{\;\;b_0}=\delta _{a_0}^{\;\;b_0}-Z_{\;\;a_1}^{b_0}\bar
D_{b_1}^{\;\;a_1}A_{a_0}^{\;\;b_1},
\end{equation}
where $\bar D_{b_1}^{\;\;a_1}$ is the inverse of $D_{b_1}^{\;\;a_1}$, the
equations (\ref{23g}) are clearly satisfied. Substituting (\ref{23k}) in (%
\ref{23h}) we find that 
\begin{equation}
\label{23l}\delta \left( {\cal P}_{a_0}-Z_{\;\;a_1}^{b_0}\bar
D_{b_1}^{\;\;a_1}A_{a_0}^{\;\;b_1}{\cal P}_{b_0}\right) =-G_{a_0}.
\end{equation}
As the co-cycles (\ref{23j}) vanish identically it results that (\ref{23h})
or (\ref{23l}) can be precisely associated with an irreducible system. In
order to derive the form of the irreducible constraints we consider the new
canonical pairs $\left( y^{a_1},\pi _{a_1}\right) $, with $\pi _{a_1}$ the
non-trivial solutions of the equations 
\begin{equation}
\label{23la}D_{a_1}^{\;\;b_1}\pi _{b_1}=\delta \left( -Z_{\;\;a_1}^{b_0}%
{\cal P}_{b_0}\right) .
\end{equation}
The equations (\ref{23la}) may have trivial, as well as non-trivial
solutions. Initially, we notice that $\pi _{b_1}=0$ (trivial solutions) if
and only if $\delta \left( -Z_{\;\;a_1}^{b_0}{\cal P}_{b_0}\right) =0$. This
case corresponds to the reducible theory with the constraints (\ref{1}) and (%
\ref{17}) (in this situation we obtain the co-cycles (\ref{23e})). The
non-trivial solutions $\pi _{b_1}\neq 0$ appear if and only if $\delta
\left( -Z_{\;\;a_1}^{b_0}{\cal P}_{b_0}\right) \neq 0$ (the quantities (\ref
{23e}) are no longer co-cycles), hence if and only if the theory is
irreducible. While within the split context the momenta $\pi _{b_1}$ replace
the dependent constraint functions, in the homological approach they enforce
the removal of the co-cycles (\ref{23e}). Expressing $\pi _{b_1}$ from (\ref
{23la}) (in the irreducible case $\pi _{b_1}\neq 0$) 
\begin{equation}
\label{23n}\pi _{b_1}=\delta \left( -Z_{\;\;a_1}^{b_0}\bar D_{b_1}^{\;\;a_1}%
{\cal P}_{b_0}\right) ,
\end{equation}
and replacing this result in (\ref{23l}) we get the relations 
\begin{equation}
\label{23m}\delta {\cal P}_{a_0}=-G_{a_0}-A_{a_0}^{\;\;b_1}\pi _{b_1}\equiv
-\gamma _{a_0}.
\end{equation}
The last formulas are nothing but the definitions of $\delta $ on the
antighost number one antighosts ${\cal P}_{a_0}$, that are attached to the
irreducible system having the constraints (\ref{21}). In conclusion, the
first-class constraints (\ref{21}) can be derived by requiring that the
non-trivial co-cycles of the type (\ref{23e}) vanish identically under the
redefinitions (\ref{23f}). For instance, in the case of free abelian
two-form gauge fields the reducible first-class constraints read as 
\begin{equation}
\label{23o1}G_i^{(2)}\equiv -2\partial ^j\pi _{ji}\approx 0.
\end{equation}
The model is first-stage redundant, namely, $Z^iG_i^{(2)}=0$, with $%
Z^i\equiv \partial ^i$. The actions of the reducible Koszul-Tate
differential on the antighost number one antighosts ${\cal P}_i$ are given
by 
\begin{equation}
\label{23o2}\delta _R{\cal P}_i=2\partial ^j\pi _{ji}.
\end{equation}
Redefining ${\cal P}_i$ such that 
\begin{equation}
\label{23o3}{\cal P}_i\rightarrow \stackrel{\sim }{\cal P}_i=\left( \delta
_i^{\;\;j}-\frac{\partial _i\partial ^j}{\triangle }\right) {\cal P}_j\equiv
D_i^{\;\;j}{\cal P}_j,
\end{equation}
from (\ref{23o2}) we find 
\begin{equation}
\label{23o4}\delta \left( {\cal P}_i-\frac{\partial _i\partial ^j}{\triangle 
}{\cal P}_j\right) =2\partial ^j\pi _{ji},
\end{equation}
where $\triangle =\partial _k\partial ^k$. Introducing the canonical pair $%
\left( \varphi ,\pi \right) $ playing the role of the variables $\left(
y^{a_1},\pi _{a_1}\right) $ and taking $D_{a_1}^{\;\;b_1}$ to be $-\triangle 
$, the equations (\ref{23la}) become 
\begin{equation}
\label{23o4a}\triangle \pi =\delta \left( \partial ^k{\cal P}_k\right) ,
\end{equation}
so 
\begin{equation}
\label{23o6}\pi =\delta \left( \frac{\partial ^k}{\triangle }{\cal P}%
_k\right) .
\end{equation}
Substituting (\ref{23o6}) in (\ref{23o4}), we find the relations 
\begin{equation}
\label{23o5}\delta {\cal P}_i=2\partial ^j\pi _{ji}+\partial _i\pi .
\end{equation}
In this way the relations (\ref{23o5}) emphasize the irreducible first-class
constraints 
\begin{equation}
\label{23o7}\gamma _i\equiv -2\partial ^j\pi _{ji}-\partial _i\pi \approx 0,
\end{equation}
which appear for instance in the example from Section 5 in the limit $M=0$
and in the absence of the fields $H^\mu $ (see the first relations in
formula (\ref{77})).

Now we can show that the constraints (\ref{1}) and (\ref{17}) are equivalent
with (\ref{21}), i.e., 
\begin{equation}
\label{24}\gamma _{a_0}\approx 0\Leftrightarrow G_{a_0}\approx 0,\;\pi
_{a_1}\approx 0. 
\end{equation}
It is simply to see that if (\ref{1}) and (\ref{17}) hold, then the
constraints (\ref{21}) also hold. The converse is valid, too. Indeed, we
will see that 
\begin{equation}
\label{25}\gamma _{a_0}\approx 0\Rightarrow G_{a_0}\approx 0,\;\pi
_{a_1}\approx 0. 
\end{equation}
This can be shown as follows. First, we apply $Z_{\;\;c_1}^{a_0}$ on (\ref
{21}), which then yields 
\begin{equation}
\label{26}\bar D_{a_1}^{\;\;c_1}Z_{\;\;c_1}^{a_0}\gamma _{a_0}=\pi _{a_1}. 
\end{equation}
With the help of (\ref{21}) and (\ref{26}) we get 
\begin{equation}
\label{27}\left( \delta _{a_0}^{\;\;b_0}-A_{a_0}^{\;\;b_1}\bar
D_{b_1}^{\;\;c_1}Z_{\;\;c_1}^{b_0}\right) \gamma _{b_0}=G_{a_0}. 
\end{equation}
From (\ref{26}-\ref{27}) we reach (\ref{25}). The Poisson brackets between
the irreducible first-class constraints read 
\begin{equation}
\label{28}\left[ \gamma _{a_0},\gamma _{b_0}\right] =\bar
C_{\;\;a_0b_0}^{c_0}\gamma _{c_0}, 
\end{equation}
where the new structure functions are expressed by 
\begin{eqnarray}\label{29}
& &\bar C_{\;\;a_0b_0}^{c_0}=C_{\;\;a_0b_0}^{d_0}\left( \delta
_{d_0}^{\;\;c_0}-A_{d_0}^{\;\;b_1}\bar
D_{b_1}^{\;\;c_1}Z_{\;\;c_1}^{c_0}\right) +\nonumber \\ 
& &\left( \left[ G_{a_0},A_{b_0}^{\;\;b_1}\right] +\left[
A_{a_0}^{\;\;b_1},G_{b_0}\right] +\left[
A_{a_0}^{\;\;c_1},A_{b_0}^{\;\;b_1}\right] \pi _{b_1}\right) \bar
D_{b_1}^{\;\;c_1}Z_{\;\;c_1}^{c_0}. 
\end{eqnarray}
The first-class Hamiltonian of the new theory can be derived starting from
the original one, $H$. Indeed, if we take 
\begin{equation}
\label{30}H^{\prime }=H+h^{a_1}\left( q^i,p_i\right) \pi _{a_1}, 
\end{equation}
with 
\begin{equation}
\label{31}\left[ h^{a_1},G_{a_0}\right] =V_{a_0}^{\;\;b_0}A_{b_0}^{\;\;a_1}, 
\end{equation}
we subsequently find 
\begin{equation}
\label{32}\left[ H^{\prime },\gamma _{a_0}\right] =\bar
V_{a_0}^{\;\;b_0}\gamma _{b_0}, 
\end{equation}
where 
\begin{equation}
\label{33}\bar V_{a_0}^{\;\;b_0}=V_{a_0}^{\;\;b_0}+\left( \left[
H,A_{a_0}^{\;\;a_1}\right] +\left[ h^{b_1},A_{a_0}^{\;\;a_1}\right] \pi
_{b_1}\right) \bar D_{a_1}^{\;\;c_1}Z_{\;\;c_1}^{b_0}. 
\end{equation}
It is clear that the first-class Hamiltonian (\ref{30}) is not unique
because we can always add to it any combinations of $\gamma _{a_0}$'s with
coefficients that are arbitrary functions. The change induced by the
modification of the Hamiltonian gives raise to a change in the structure
functions (\ref{33}).

In brief, in this section we constructed an irreducible first-class system
associated with the original redundant one, described by the constraints (%
\ref{21}) and the first-class Hamiltonian (\ref{30}), displaying the gauge
algebra (\ref{28}) and (\ref{32}). The irreducible theory built here will be
important by virtue of the subsequent development.

\section{Irreducible quantization of the reducible theory}

Now, we show that the reducible, respectively, irreducible theories possess
the same classical observables. We start from an observable $F$ of the
irreducible theory. Accordingly, $F$ should verify the equations 
\begin{equation}
\label{34}\left[ F,\gamma _{a_0}\right] \approx 0. 
\end{equation}
On account of (\ref{21}) and (\ref{26}), from (\ref{34}) we deduce 
\begin{equation}
\label{35}\left[ F,G_{a_0}\right] +\left[ F,\pi _{a_1}\right]
A_{a_0}^{\;\;a_1}\approx 0. 
\end{equation}
On the other hand, multiplying (\ref{35}) with $Z_{\;\;b_1}^{a_0}$ and using
(\ref{27}), we arrive at 
\begin{equation}
\label{36}\left[ F,\pi _{a_1}\right] D_{b_1}^{\;\;a_1}\approx \left[
F,Z_{\;\;b_1}^{a_0}\right] \left( \delta
_{a_0}^{\;\;b_0}-A_{a_0}^{\;\;c_1}\bar
D_{c_1}^{\;\;d_1}Z_{\;\;d_1}^{b_0}\right) \gamma _{b_0}\approx 0. 
\end{equation}
Because $D_{b_1}^{\;\;a_1}$ has maximal rank (see (\ref{23})), from (\ref{36}%
) we infer 
\begin{equation}
\label{37}\left[ F,\pi _{a_1}\right] \approx 0, 
\end{equation}
such that 
\begin{equation}
\label{38}\left[ F,\gamma _{a_0}\right] \approx 0\Rightarrow \left[
F,G_{a_0}\right] \approx 0. 
\end{equation}
In conclusion, if $F$ is an observable of the irreducible theory, then it is
also an observable of the original reducible one. The converse is valid,
too, because any observable of the reducible theory verifies the equations $%
\left[ F,G_{a_0}\right] \approx 0$ and does not depend on the newly added
canonical variables, such that (\ref{34}) are indeed satisfied. Thus, both
the irreducible and reducible models display the same physical observables.
A simple count shows that the numbers of physical degrees of freedom of the
reducible, respectively, irreducible theories are both equal to $N-M_0+M_1$.
The last conclusions prove that the original reducible theory is physically
equivalent with the irreducible one. This makes permissible the replacement
of the BRST quantization for the original redundant system by the BRST
quantization of the irreducible theory.

The first attempt at quantizing the irreducible system is to apply the
antifield-BRST formalism with respect to its extended action, namely, 
\begin{equation}
\label{39}S_0^{\prime E}\left[ q^i,p_i,y^{a_1},\pi _{a_1},u^{a_0}\right]
=\int dt\left( \dot q^ip_i+\dot y^{a_1}\pi _{a_1}-H^{\prime }-u^{a_0}\gamma
_{a_0}\right) . 
\end{equation}
Action (\ref{39}) is invariant under the gauge transformations 
\begin{equation}
\label{40}\delta _\epsilon F=\left[ F,\gamma _{a_0}\right] \epsilon
^{a_0},\;\delta _\epsilon u^{a_0}=\dot \epsilon ^{a_0}-\bar
V_{b_0}^{\;\;a_0}\epsilon ^{b_0}-\bar C_{\;\;b_0c_0}^{a_0}u^{b_0}\epsilon
^{c_0}, 
\end{equation}
with $\epsilon ^{a_0}$ the gauge parameters associated with the irreducible
constraints (\ref{21}). In the absence of the newly introduced variables,
the extended action (\ref{39}) together with its gauge transformations, (\ref
{40}), should reduce to those from the reducible case. The gauge
transformations of the Lagrange multipliers from (\ref{40}) do not lead to
the corresponding transformations from the reducible situation because the
terms $-Z_{\;\;a_1}^{a_0}\epsilon ^{a_1}$ are missing. The gauge parameters $%
\epsilon ^{a_1}$, which were attached to the first-stage reducibility
functions, are absent within the irreducible approach. In order to restore
these terms, it is necessary to further enlarge the phase-space by adding
some supplementary canonical pairs $\left( z_1^{a_1},p_{1a_1}\right) $, $%
\left( z_2^{a_1},p_{2a_1}\right) $, subject to the constraints 
\begin{equation}
\label{41}-p_{1a_1}\approx 0,\;p_{2a_1}\approx 0. 
\end{equation}
Obviously, (\ref{21}) and (\ref{41}) are still first-class and irreducible.
Adding to the first set of constraints from (\ref{41}) a combination of
first-class constraints (see (\ref{26})), we obtain the equivalent
first-class set 
\begin{equation}
\label{42}\gamma _{a_1}\equiv \pi _{a_1}-p_{1a_1}\approx 0,\;\gamma
_{a_1}^{\prime }\equiv p_{2a_1}\approx 0. 
\end{equation}
The additional first-class constraints do not afflict the number of physical
degrees of freedom of the former irreducible system. At the same time, the
above established equivalence between the physical observables respectively
associated with the reducible and irreducible theories remains valid. This
is because an observable $F$ of the last irreducible model must check,
beside (\ref{34}), the equations 
\begin{equation}
\label{43}\left[ F,\gamma _{a_1}\right] \approx 0,\;\left[ F,\gamma
_{a_1}^{\prime }\right] \approx 0. 
\end{equation}
On account of (\ref{37}) relations (\ref{43}) indicate that in addition to
the prior conditions $F$ does not depend (at least weakly) on the last added
canonical pairs. As it will be seen below, the constraints (\ref{42}) will
imply the presence of the terms $-Z_{\;\;a_1}^{a_0}\epsilon ^{a_1}$ within
the gauge transformations of the Lagrange multipliers $u^{a_0}$. In this
sense, the constraints (\ref{42}) play in a certain way the role of the
original reducibility relations. The first-class Hamiltonian with respect to
the first-class set (\ref{21}) and (\ref{42}) can be taken as 
\begin{equation}
\label{44}H_0=H^{\prime }+z_2^{a_1}Z_{\;\;a_1}^{a_0}\gamma
_{a_0}+y^{a_1}\gamma _{a_1}^{\prime }, 
\end{equation}
such that the new irreducible gauge algebra reads 
\begin{equation}
\label{45}\left[ \gamma _{a_0},\gamma _{b_0}\right] =\bar
C_{\;\;a_0b_0}^{c_0}\gamma _{c_0},\;\left[ \gamma _{a_0},\gamma
_{a_1}\right] =0,\;\left[ \gamma _{a_0},\gamma _{a_1}^{\prime }\right] =0, 
\end{equation}
\begin{equation}
\label{46}\left[ \gamma _{a_1},\gamma _{b_1}\right] =0,\;\left[ \gamma
_{a_1},\gamma _{b_1}^{\prime }\right] =0,\;\left[ \gamma _{a_1}^{\prime
},\gamma _{b_1}^{\prime }\right] =0, 
\end{equation}
\begin{equation}
\label{47}\left[ H_0,\gamma _{a_0}\right] =\tilde V_{a_0}^{\;\;b_0}\gamma
_{b_0}+A_{a_0}^{\;\;b_1}\gamma _{b_1}^{\prime },\;\left[ H_0,\gamma
_{a_1}\right] =\gamma _{a_1}^{\prime },\;\left[ H_0,\gamma _{a_1}^{\prime
}\right] =Z_{\;\;a_1}^{a_0}\gamma _{a_0}. 
\end{equation}
Simple calculations show that the functions $\tilde V_{a_0}^{\;\;b_0}$ from (%
\ref{47}) are of the form 
\begin{equation}
\label{48}\tilde V_{a_0}^{\;\;b_0}=\bar V_{a_0}^{\;\;b_0}+z_2^{a_1}\left(
\mu _{\;\;a_0a_1}^{b_0c_0}G_{c_0}+\lambda
_{\;\;a_0a_1}^{b_1}Z_{\;\;b_1}^{b_0}+\left[
Z_{\;\;a_1}^{b_0},A_{a_0}^{\;\;b_1}\right] \pi _{b_1}\right) , 
\end{equation}
with $\mu _{\;\;a_0a_1}^{b_0c_0}$ and $\lambda _{\;\;a_0a_1}^{b_1}$
appearing in 
\begin{equation}
\label{49}\left[ Z_{\;\;a_1}^{c_0},G_{d_0}\right]
=-Z_{\;\;a_1}^{a_0}C_{\;\;a_0d_0}^{c_0}+\mu
_{\;\;d_0a_1}^{c_0b_0}G_{b_0}+\lambda _{\;\;d_0a_1}^{b_1}Z_{\;\;b_1}^{c_0}, 
\end{equation}
and $\mu _{\;\;a_0a_1}^{b_0c_0}$ antisymmetric in the upper indices, i.e., $%
\mu _{\;\;a_0a_1}^{b_0c_0}=-\mu _{\;\;a_0a_1}^{c_0b_0}$. The relations (\ref
{49}) can be inferred by taking the Poisson brackets between (\ref{2}) and $%
G_{d_0}$, which leads to 
\begin{equation}
\label{50}\left( Z_{\;\;a_1}^{a_0}C_{\;\;a_0d_0}^{c_0}+\left[
Z_{\;\;a_1}^{c_0},G_{d_0}\right] \right) G_{c_0}=0. 
\end{equation}
From (\ref{50}) and (\ref{2}) it follows directly (\ref{49}). We outline
that the functions $\bar C_{\;\;a_0b_0}^{c_0}$ and $\tilde V_{a_0}^{\;\;b_0}$
encode the reducible structure within the irreducible theory. The
first-class Hamiltonian (\ref{44}) is unique up to a combination in terms of
the functions $\gamma _{a_0}$, $\gamma _{a_1}$ and $\gamma _{a_1}^{\prime }$%
. The change of $H_0$ consequently implies the change of the structure
functions from (\ref{47}).

The extended action describing the new irreducible theory%
\begin{eqnarray}\label{51}
& &S_0^E\left[ q^i,p_i,y^{a_1},\pi
_{a_1},z_1^{a_1},p_{1a_1},z_2^{a_1},p_{2a_1},u^{a_0},u^{a_1},v^{a_1}\right]
=\int dt\left( \dot q^ip_i+\right. \nonumber \\
& &\left. \dot y^{a_1}\pi _{a_1}+\dot
z_1^{a_1}p_{1a_1}+\dot z_2^{a_1}p_{2a_1}-H_0-u^{a_0}\gamma
_{a_0}-u^{a_1}\gamma _{a_1}-v^{a_1}\gamma _{a_1}^{\prime }\right) , 
\end{eqnarray}
is invariant under the gauge transformations 
\begin{equation}
\label{52}\delta _\epsilon F=\left[ F,\gamma _{a_0}\right] \epsilon
^{a_0}+\left[ F,\gamma _{a_1}\right] \epsilon _1^{a_1}+\left[ F,\gamma
_{a_1}^{\prime }\right] \epsilon _2^{a_1}, 
\end{equation}
\begin{equation}
\label{53}\delta _\epsilon u^{a_0}=\dot \epsilon ^{a_0}-\tilde
V_{b_0}^{\;\;a_0}\epsilon ^{b_0}-\bar C_{\;\;b_0c_0}^{a_0}u^{b_0}\epsilon
^{c_0}-Z_{\;\;a_1}^{a_0}\epsilon _2^{a_1}, 
\end{equation}
\begin{equation}
\label{54}\delta _\epsilon u^{a_1}=\dot \epsilon _1^{a_1},\;\delta _\epsilon
v^{a_1}=\dot \epsilon _2^{a_1}-A_{a_0}^{\;\;a_1}\epsilon ^{a_0}-\epsilon
_1^{a_1}. 
\end{equation}
We emphasize that in this way the terms $-Z_{\;\;a_1}^{a_0}\epsilon _2^{a_1}$
are restored within the gauge transformations of the multipliers $u^{a_0}$.
This is precisely the effect of introducing the supplementary pairs $\left(
z_1^{a_1},p_{1a_1}\right) $, $\left( z_2^{a_1},p_{2a_1}\right) $ subject to
the constraints (\ref{42}). If in (\ref{51}-\ref{54}) we discard all the
newly introduced canonical pairs, we get the extended action and the gauge
transformations from the initial redundant case.

With these elements at hand, it appears clearly that we can substitute the
quantization of the initial theory by the quantization of the last
irreducible system. In the sequel we perform the antifield-BRST quantization
with respect to the action (\ref{51}). To this end, we introduce the ghosts 
\begin{equation}
\label{55}\left( \eta ^{a_0},\eta _1^{a_1},\eta _2^{a_1}\right) , 
\end{equation}
and also the antifields 
\begin{equation}
\label{56}\left( q_i^{*},p^{*i},y_{a_1}^{*},\pi
^{*a_1},z_{1a_1}^{*},p_1^{*a_1},z_{2a_1}^{*},p_2^{*a_1},u_{a_0}^{*},
u_{a_1}^{*},v_{a_1}^{*},\eta _{a_0}^{*},\eta _{1a_1}^{*}, \eta
_{2a_1}^{*}\right) . 
\end{equation}
The ghosts have ghost number one, the antifields associated with the
variables involved with (\ref{51}) possess ghost number minus one, while the
antifields of the ghosts have ghost number minus two. The solution to the
master equation is given by 
\begin{eqnarray}\label{57}
& &S^E=S_0^E+\int dt\left( q_i^{*}\left[ q^i,\gamma _{a_0}\right] \eta
^{a_0}+p^{*i}\left[ p_i,\gamma _{a_0}\right] \eta ^{a_0}+\right. 
\nonumber \\ 
& &y_{a_1}^{*}\left( A_{a_0}^{\;\;a_1}\eta ^{a_0}+\eta _1^{a_1}\right)
-z_{1a_1}^{*}\eta _1^{a_1}+z_{2a_1}^{*}\eta _2^{a_1}+\nonumber \\ 
& &u_{a_0}^{*}\left( \dot \eta ^{a_0}-\tilde 
V_{b_0}^{\;\;a_0}\eta ^{b_0}-\bar
C_{\;\;b_0c_0}^{a_0}u^{b_0}\eta ^{c_0}-Z_{\;\;a_1}^{a_0}\eta _2^{a_1}\right)
+\nonumber \\ 
& &\left. u_{a_1}^{*}\dot \eta _1^{a_1}+v_{a_1}^{*}\left( \dot \eta
_2^{a_1}-A_{a_0}^{\;\;a_1}\eta ^{a_0}-\eta _1^{a_1}\right) -\frac 12\eta
_{a_0}^{*}\bar C_{\;\;b_0c_0}^{a_0}\eta ^{b_0}\eta ^{c_0}+\cdots \right) . 
\end{eqnarray}
In order to derive a gauge-fixed action, it is necessary to fix the gauge.
In this respect, it is useful to take a gauge-fixing fermion 
\begin{equation}
\label{58}\psi =\psi \left[ q^i,p_i,y^{a_1},\pi
_{a_1},z_1^{a_1},p_{1a_1},z_2^{a_1},p_{2a_1},\eta ^{a_0},\eta _1^{a_1},\eta
_2^{a_1},u_{a_0}^{*},u_{a_1}^{*},v_{a_1}^{*}\right] , 
\end{equation}
implementing some irreducible gauge conditions, with the help of which we
eliminate all the antifields excepting $u_{a_0}^{*}$, $u_{a_1}^{*}$, $%
v_{a_1}^{*}$, that are maintained in favour of their fields. The possibility
to build some irreducible gauge conditions is easier on behalf of the newly
added canonical pairs, which play at this level the same role like the
auxiliary variables from the reducible approach. We can put the gauge-fixed
action under a form displaying a more direct link with the Hamiltonian BRST
quantization of the irreducible system following the procedure exposed in 
\cite{12}--\cite{13}. In this light, we declare the variables $\left( \eta
^{a_0},u_{a_0}^{*}\right) $, $\left( \eta _1^{a_1},u_{a_1}^{*}\right) $, $%
\left( \eta _2^{a_1},v_{a_1}^{*}\right) $ respectively conjugated in the
Poisson bracket 
\begin{equation}
\label{59}\left[ u_{a_0}^{*},\eta ^{b_0}\right] =-\delta
_{a_0}^{\;\;b_0},\;\left[ u_{a_1}^{*},\eta _1^{b_1}\right] =-\delta
_{a_1}^{\;\;b_1},\;\left[ v_{a_1}^{*},\eta _2^{b_1}\right] =-\delta
_{a_1}^{\;\;b_1}, 
\end{equation}
and regard the antifields like the momenta associated with the ghosts. Under
these circumstances, the gauge-fixed action corresponding to (\ref{57})
reads as%
\begin{eqnarray}\label{60}
& &S_\psi =\int dt\left( \dot q^ip_i+\dot y^{a_1}\pi _{a_1}+\dot
z_1^{a_1}p_{1a_1}+\dot z_2^{a_1}p_{2a_1}+\right. \nonumber \\ 
& &\left. u_{a_0}^{*}\dot \eta ^{a_0}+u_{a_1}^{*}\dot \eta
_1^{a_1}+v_{a_1}^{*}\dot \eta _2^{a_1}-H_B+\left[ \psi ,\Omega \right]
\right) , 
\end{eqnarray}
where the BRST charge, respectively, the BRST-extension of the first-class
Hamiltonian start like 
\begin{equation}
\label{61}\Omega =\gamma _{a_0}\eta ^{a_0}+\gamma _{a_1}\eta _1^{a_1}+\gamma
_{a_1}^{\prime }\eta _2^{a_1}+\frac 12u_{a_0}^{*}\bar
C_{\;\;b_0c_0}^{a_0}\eta ^{b_0}\eta ^{c_0}+\cdots , 
\end{equation}
\begin{equation}
\label{62}H_B=H_0+u_{a_0}^{*}\left( \tilde V_{b_0}^{\;\;a_0}\eta
^{b_0}+Z_{\;\;a_1}^{a_0}\eta _2^{a_1}\right) +v_{a_1}^{*}\left(
A_{a_0}^{\;\;a_1}\eta ^{a_0}+\eta _1^{a_1}\right) +\cdots . 
\end{equation}
This completes our irreducible procedure in the case of first-stage
reducible first-class Hamiltonian theories. Until now, we showed how a
first-stage reducible first-class Hamiltonian system can be quantized in the
framework of the irreducible antifield-BRST formalism, i.e., without
introducing ghosts of ghosts.

\section{$L$-stage reducible Hamiltonian theories}

In this section we generalize the results from the first-stage case to
higher-order-stage reducible systems. If the original Hamiltonian theory is $%
L$-stage reducible (with finite $L$), the construction of the corresponding
irreducible system goes along the same line like that from the first-stage
case. We assume the reducibility relations 
\begin{equation}
\label{63}Z_{\;\;a_1}^{a_0}G_{a_0}=0,\;Z_{\;\;a_1}^{a_0}Z_{\;\;a_2}^{a_1}=0,%
\cdots ,Z_{\;\;a_{L-1}}^{a_{L-2}}Z_{\;\;a_L}^{a_{L-1}}=0, 
\end{equation}
with $a_k=1,\cdots ,M_k$. Next, we introduce the canonical pairs $\left(
y^{a_k},\pi _{a_k}\right) _{k=1,\cdots ,L}$ corresponding to the free
indices of the above reducibility relations, and constrain these new
variables like 
\begin{equation}
\label{63a}\pi _{a_k}\approx 0. 
\end{equation}
Constraints (\ref{1}) and (\ref{63a}) are first-class and obviously
reducible. In a manner similar with that from section 2, we derive the
first-class constraints 
\begin{equation}
\label{64}\gamma _{a_0}\equiv G_{a_0}+A_{a_0}^{\,\;\;b_1}\pi _{b_1}\approx
0, 
\end{equation}
\begin{equation}
\label{65}\gamma _{a_{2k}}\equiv Z_{\;\;a_{2k}}^{a_{2k-1}}\pi
_{a_{2k-1}}+A_{a_{2k}}^{\;\;a_{2k+1}}\pi _{a_{2k+1}}\approx 0,\;k=1,\cdots
,a, 
\end{equation}
\begin{equation}
\label{66}\bar \gamma _{a_{2k}}\equiv \pi _{a_{2k}}\approx 0,\;k=1,\cdots
,a, 
\end{equation}
which are equivalent with (\ref{1}) and (\ref{63a}). Acting like in the
first-stage situation, we find that 
\begin{equation}
\label{66a}\pi _{a_{2k+1}}=m_{a_{2k+1}}^{\;\;b_{2k}}\gamma
_{b_{2k}},\;G_{a_0}=m_{a_0}^{\;\;b_{2k}}\gamma _{b_{2k}},\;k=0,\cdots ,b, 
\end{equation}
for some appropriate functions $m_{a_{2k+1}}^{\;\;b_{2k}}$ and $%
m_{a_0}^{\;\;b_{2k}}$, such that the equivalence between (\ref{1}), (\ref
{63a}) and (\ref{64}--\ref{66}) is direct. We employed the notations 
\begin{equation}
\label{68a}a=\left\{ 
\begin{array}{c}
\frac L2,\; 
{\rm for}\;L\;{\rm even}, \\ \frac{L-1}2,\;{\rm for}\;L\;{\rm odd}, 
\end{array}
\right. 
\end{equation}
\begin{equation}
\label{68}b=\left\{ 
\begin{array}{c}
\frac L2-1,\; 
{\rm for}\;L\;{\rm even}, \\ \frac{L-1}2,\;{\rm for}\;L\;{\rm odd}. 
\end{array}
\right. 
\end{equation}
In (\ref{65}) the functions $A_{a_k}^{\;\;a_{k+1}}$ depend only on $\left(
q^i,p_i\right) $ and possess the property 
\begin{equation}
\label{67}rank\left( Z_{\;\;a_k}^{a_{k-1}}A_{a_{k-1}}^{\;\;b_k}\right)
\approx \sum\limits_{i=k}^L\left( -\right) ^{k+i}M_i. 
\end{equation}
Moreover, the $A_{a_{k-1}}^{\;\;b_k}$'s can be taken to satisfy the
relations 
\begin{equation}
\label{67a}A_{a_{k-1}}^{\;\;b_k}A_{b_k}^{\;\;a_{k+1}}=0. 
\end{equation}
The last relations are based on the fact that we can always choose the $%
A_{a_{k-1}}^{\;\;b_k}$'s proportional with the transposed of $%
Z_{\;\;b_k}^{a_{k-1}}$'s. On account of (\ref{67a}), one finds that the
first-class set (\ref{64}--\ref{66}) is irreducible. We remark that (\ref{66}%
) are irreducible. Thus, it remains to be proved that (\ref{64}--\ref{65})
are so. This can be seen by multiplying (\ref{64}) with $Z_{\;\;b_1}^{a_0}$
and (\ref{65}) with $Z_{\;\;b_{2k+1}}^{a_{2k}}$, which induce 
\begin{equation}
\label{67b}Z_{\;\;b_1}^{a_0}\gamma
_{a_0}=Z_{\;\;b_1}^{a_0}A_{a_0}^{\,\;\;c_1}\pi
_{c_1},\;Z_{\;\;b_{2k+1}}^{a_{2k}}\gamma
_{a_{2k}}=Z_{\;\;b_{2k+1}}^{a_{2k}}A_{a_{2k}}^{\;\;a_{2k+1}}\pi _{a_{2k+1}}. 
\end{equation}
With the help of (\ref{67b}) and (\ref{67a}) we infer that $%
Z_{\;\;b_1}^{a_0}\gamma _{a_0}=0$, $Z_{\;\;b_{2k+1}}^{a_{2k}}\gamma
_{a_{2k}}=0$ if and only if 
\begin{equation}
\label{67c}\pi _{a_{2k+1}}=A_{a_{2k+1}}^{\,\;\;a_{2k+2}}\nu
_{a_{2k+2}},\;k=0,\cdots ,b, 
\end{equation}
with $\nu _{a_{2k+2}}$ some functions. Replacing (\ref{67c}) in (\ref{64}--%
\ref{65}) we obtain 
\begin{equation}
\label{67d}G_{a_0}\approx
0,\;Z_{\;\;b_{2k}}^{a_{2k-1}}A_{a_{2k-1}}^{\;\;a_{2k}}\nu _{a_{2k}}\approx
0, 
\end{equation}
which leads, by virtue of (\ref{67}--\ref{67a}), to 
\begin{equation}
\label{67e}\nu _{a_{2k}}\approx A_{a_{2k}}^{\;\;a_{2k+1}}\lambda
_{a_{2k+1}}, 
\end{equation}
for some $\lambda _{a_{2k+1}}$. Substituting (\ref{67e}) in (\ref{67c}) we
derive that (\ref{64}--\ref{65}) are reducible with the reducibility
functions $Z_{\;\;b_{2k+1}}^{a_{2k}}$ if and only if $\pi _{a_{2k+1}}\approx
0$. In this situation the constraints (\ref{64}--\ref{65}) and (\ref{66})
are nothing but (\ref{1}) and (\ref{63a}). Thus, (\ref{64}--\ref{66}) are
reducible with the reducibility functions $Z_{\;\;b_{2k+1}}^{a_{2k}}$ if and
only if they have the form (\ref{1}) and (\ref{63a}). On the other hand, if
one multiplies (\ref{65}) by $A_{b_{2k-1}}^{\;\;a_{2k}}$, one gets 
\begin{equation}
\label{67f}A_{b_{2k-1}}^{\;\;a_{2k}}\gamma
_{a_{2k}}=A_{b_{2k-1}}^{\;\;a_{2k}}Z_{\;\;a_{2k}}^{a_{2k-1}}\pi _{a_{2k-1}}, 
\end{equation}
due to (\ref{67a}). From (\ref{67f}), it results that $A_{b_{2k-1}}^{\;%
\;a_{2k}}\gamma _{a_{2k}}=0$ if and only if 
\begin{equation}
\label{67g}\pi _{a_{2k-1}}=Z_{\;\;a_{2k-1}}^{a_{2k-2}}\mu _{a_{2k-2}}, 
\end{equation}
for some $\mu _{a_{2k-2}}$. Inserting (\ref{67g}) in (\ref{65}) we find 
\begin{equation}
\label{67h}A_{a_{2k}}^{\;\;a_{2k+1}}Z_{\;\;a_{2k+1}}^{b_{2k}}\mu
_{b_{2k}}\approx 0, 
\end{equation}
which leads to 
\begin{equation}
\label{67i}\mu _{b_{2k}}\approx Z_{\;\;b_{2k}}^{a_{2k-1}}\sigma _{a_{2k-1}}, 
\end{equation}
for some $\sigma _{a_{2k-1}}$. Introducing (\ref{67i}) in (\ref{67g}) we
conclude that (\ref{64}--\ref{66}) are reducible with the reducibility
functions $A_{b_{2k-1}}^{\;\;a_{2k}}$ if and only if they reduce to (\ref{1}%
) and (\ref{63a}). In this way, the irreducibility of (\ref{64}--\ref{66})
is completely proved.

In the meantime, it is still necessary to add the pairs $\left(
z_1^{a_{2k+1}},p_{1a_{2k+1}}\right) $, $\left(
z_2^{a_{2k+1}},p_{2a_{2k+1}}\right) $, with $k=0,\cdots ,b$. With the help
of the last pairs we build the supplementary first-class constraints 
\begin{equation}
\label{69}\gamma _{a_{2k+1}}\equiv \pi _{a_{2k+1}}-p_{1a_{2k+1}}\approx
0,\;\gamma _{a_{2k+1}}^{\prime }\equiv p_{2a_{2k+1}}\approx 0. 
\end{equation}
The equivalence between the observables of the original redundant theory and
those of the irreducible system is gained like in the first-stage situation.
We illustrate the proof of the equivalence in the case $L$ odd, the other
situation being treated in a similar fashion. If $F$ stands for an
observable of the irreducible system, the conditions $\left[ F,\bar \gamma
_{a_{2k}}\right] \approx 0$ indicate that it does not depend, at least
weakly, on $y^{a_{2k}}$. In the meantime, $F$ should verify 
\begin{equation}
\label{69a}\left[ F,\gamma _{a_{2k}}\right] \approx 0,\;k=0,\cdots ,a. 
\end{equation}
We start from the last relation (\ref{69a}) (for $k=a$). On account of (\ref
{66a}), we obtain 
\begin{equation}
\label{69b}\left[ F,\pi _{a_{L-2}}\right] Z_{\;\;a_{L-1}}^{a_{L-2}}+\left[
F,\pi _{a_L}\right] A_{a_{L-1}}^{\;\;a_L}\approx 0. 
\end{equation}
Multiplying the above equation by $Z_{\;\;b_L}^{a_{L-1}}$, on behalf of (\ref
{67}), and as $M_{L+1}=0$, we derive 
\begin{equation}
\label{69c}\left[ F,\pi _{a_L}\right] \approx 0, 
\end{equation}
such that (\ref{69b}) becomes 
\begin{equation}
\label{69d}\left[ F,\pi _{a_{L-2}}\right] Z_{\;\;a_{L-1}}^{a_{L-2}}\approx
0. 
\end{equation}
Multiplying the next equation from (\ref{69a}) (for $k=a-1$) with $%
Z_{\;\;b_{L-2}}^{a_{L-3}}$, we further infer 
\begin{equation}
\label{69e}\left[ F,\pi _{a_{L-2}}\right]
A_{a_{L-3}}^{\;\;a_{L-2}}Z_{\;\;b_{L-2}}^{a_{L-3}}\approx 0. 
\end{equation}
With the help of (\ref{67a}), from (\ref{69e}) we deduce 
\begin{equation}
\label{69f}\left[ F,\pi _{a_{L-2}}\right]
=n_{b_{L-1}}A_{a_{L-2}}^{\;\;b_{L-1}}, 
\end{equation}
for some functions $n_{b_{L-1}}$. Replacing (\ref{69f}) in (\ref{69d}) it
follows that the weak relations $n_{b_{L-1}}A_{a_{L-2}}^{\;\;b_{L-1}}Z_{\;%
\;a_{L-1}}^{a_{L-2}}\approx 0$ imply 
\begin{equation}
\label{69g}n_{b_{L-1}}\approx \rho _{b_L}A_{b_{L-1}}^{\;\;b_L}, 
\end{equation}
for some $\rho _{b_L}$. Inserting (\ref{69g}) in (\ref{69f}) we get 
\begin{equation}
\label{69h}\left[ F,\pi _{a_{L-2}}\right] \approx 0, 
\end{equation}
due to (\ref{67a}). Reprising the same steps on the remaining equations (\ref
{69a}) we consequently arrive to 
\begin{equation}
\label{69i}\left[ F,\pi _{a_{L-2k}}\right] \approx 0, 
\end{equation}
which lead to 
\begin{equation}
\label{69j}\left[ F,G_{a_0}\right] \approx 0. 
\end{equation}
Moreover, the equations $\left[ F,\gamma _{a_{2k+1}}\right] \approx 0$ and $%
\left[ F,\gamma _{a_{2k+1}}^{\prime }\right] \approx 0$ express the fact
that $F$ does not depend on the $z$'s. Thus, any observable of the
irreducible theory does not involve, at least weakly, the newly introduced
variables, and, in addition, it satisfies (\ref{69j}), which are nothing but
the equations that should be checked by any observable of the original
redundant system, which show that $F$ is also an observable of the reducible
theory. Conversely, if $F^{\prime }$ denotes an observable associated with
the reducible system, then it is obviously an observable of the irreducible
theory.

The first-class Hamiltonian with respect to the irreducible first-class
constraints (\ref{64}--\ref{66}) and (\ref{69}) can be taken under the form%
\begin{eqnarray}\label{69k}
& &H_0^{\prime }=H^{\prime }+\sum\limits_{k=1}^ay^{a_{2k}}\gamma
_{a_{2k}}+\sum\limits_{k=0}^by^{a_{2k+1}}p_{2a_{2k+1}}+\nonumber \\ 
& &\sum\limits_{k=0}^bz_2^{a_{2k+1}}\left(
Z_{\;\;a_{2k+1}}^{a_{2k}}\gamma _{a_{2k}}+A_{a_{2k+1}}^{\;\;a_{2k+2}}\gamma
_{a_{2k+2}}\right) , 
\end{eqnarray}
with $H^{\prime }$ given by (\ref{30}), where we understood the convention $%
f^{a_k}=0$ if $k<0$ or $k>L$. The first-class Hamiltonian (\ref{69k}) is
again unique up to adding a combination in the first-class constraint
functions.

With all these elements at hand, the quantization of the irreducible theory
goes from now on along the standard antifield-BRST rules. The ghost spectrum
contains only the ghost number one variables associated with the
corresponding constraint functions 
\begin{equation}
\label{69la}\eta ^{a_0}\leftrightarrow \gamma _{a_0},\;\eta
_1^{a_{2k}}\leftrightarrow \bar \gamma _{a_{2k}},\;\eta
_2^{a_{2k}}\leftrightarrow \gamma _{a_{2k}},\;k=1,\cdots ,a, 
\end{equation}
\begin{equation}
\label{69l}\eta _1^{a_{2k+1}}\leftrightarrow \gamma _{a_{2k+1}},\;\eta
_2^{a_{2k+1}}\leftrightarrow \gamma _{a_{2k+1}}^{\prime },\;k=0,\cdots ,b, 
\end{equation}
while the antifield sector is given by 
\begin{equation}
\label{69ma}\left( q_i^{*},p^{*i}\right) ,\left( y_{a_k}^{*},\pi
^{*a_k}\right) _{k=1,\cdots ,L},\left(
z_{1a_{2k+1}}^{*},p_1^{*a_{2k+1}}\right) _{k=0,\cdots ,b},\left(
z_{2a_{2k+1}}^{*},p_2^{*a_{2k+1}}\right) _{k=0,\cdots ,b}, 
\end{equation}
\begin{equation}
\label{69mb}u_{a_0}^{*},\left( u_{a_{2k}}^{*},v_{a_{2k}}^{*}\right)
_{k=1,\cdots ,a},\left( u_{a_{2k+1}}^{*},v_{a_{2k+1}}^{*}\right)
_{k=0,\cdots ,b}, 
\end{equation}
\begin{equation}
\label{69m}\eta _{a_0}^{*},\left( \eta _{1a_{2k}}^{*},\eta
_{2a_{2k}}^{*}\right) _{k=1,\cdots ,a},\left( \eta _{1a_{2k+1}}^{*},\eta
_{2a_{2k+1}}^{*}\right) _{k=0,\cdots ,b}. 
\end{equation}
The antifields $\left( u_{a_{2k}}^{*},v_{a_{2k}}^{*}\right) $ correspond to
the Lagrange multipliers of the constraint functions $\bar \gamma _{a_{2k}}$%
, respectively, $\gamma _{a_{2k}}$, and $\left(
u_{a_{2k+1}}^{*},v_{a_{2k+1}}^{*}\right) $ are associated with $\gamma
_{a_{2k+1}}$, respectively, $\gamma _{a_{2k+1}}^{\prime }$. The variables (%
\ref{69ma}-\ref{69mb}) have ghost number minus one, while in (\ref{69m})
there appear only ghost number minus two antifields. The gauge-fixing
fermion should be taken to depend on the $\Phi ^A$'s, on the ghosts, and
also on the antifields of the Lagrange multipliers, where 
\begin{equation}
\label{69mc}\Phi ^A=\left( q^i,p_i,y^{a_k},\pi
_{a_k},z_1^{a_{2k+1}},p_{1a_{2k+1}},z_2^{a_{2k+1}},p_{2a_{2k+1}}\right) . 
\end{equation}
With the help of the gauge-fixing fermion we eliminate all the antifields
except the antifields of the multipliers, and also the Lagrange multipliers.
The gauge-fixed action will be expressed by 
\begin{eqnarray}\label{69n}
& &S_\psi =\int dt\left( \dot q^ip_i+\sum\limits_{k=1}^L\dot y^{a_k}\pi
_{a_k}+\sum\limits_{k=0}^b\left( \dot z_1^{a_{2k+1}}p_{1a_{2k+1}}+\dot
z_2^{a_{2k+1}}p_{2a_{2k+1}}\right) +\right. \nonumber \\ 
& &\left. u_{a_0}^{*}\dot \eta ^{a_0}+\sum\limits_{k=1}^L\left(
u_{a_k}^{*}\dot \eta _1^{a_k}+v_{a_k}^{*}\dot \eta _2^{a_k}\right)
-H_B+\left[ \psi ,\Omega \right] \right) , 
\end{eqnarray}
where the BRST charge and the BRST extension of the first-class Hamiltonian (%
\ref{69k}) respectively start like 
\begin{eqnarray}\label{69o}
& &\Omega =\gamma _{a_0}\eta ^{a_0}+\sum\limits_{k=1}^a\left( \bar \gamma
_{a_{2k}}\eta _1^{a_{2k}}+\gamma _{a_{2k}}\eta _2^{a_{2k}}\right) + 
\nonumber \\
& &\sum\limits_{k=0}^b\left( \gamma _{a_{2k+1}}\eta _1^{a_{2k+1}}+\gamma
_{a_{2k+1}}^{\prime }\eta _2^{a_{2k+1}}\right) +\frac 12u_{a_0}^{*}\bar
C_{\;\;b_0c_0}^{a_0}\eta ^{b_0}\eta ^{c_0}+\nonumber \\ 
& &\frac
12\sum\limits_{j=1}^a\sum\limits_{k=1}^a\sum\limits_{i=1}^av_{a_{2j}}^{*}%
\bar C_{\;\;b_{2k}c_{2i}}^{a_{2j}}\eta _2^{b_{2k}}\eta _2^{c_{2i}}+\cdots , 
\end{eqnarray}
\begin{eqnarray}\label{69p}
& &H_B=H_0^{\prime }+u_{a_0}^{*}\left( \tilde V_{b_0}^{\;\;a_0}\eta
^{b_0}+Z_{\;\;a_1}^{a_0}\eta _2^{a_1}\right) +\nonumber \\ 
& &\sum\limits_{k=1}^av_{a_{2k}}^{*}\left( \eta
_1^{a_{2k}}+\sum\limits_{j=1}^a\tilde V_{b_{2j}}^{\;\;a_{2k}}\eta
_2^{b_{2j}}+Z_{\;\;a_{2k+1}}^{a_{2k}}\eta
_2^{a_{2k+1}}+A_{a_{2k-1}}^{\;\;a_{2k}}\eta _2^{a_{2k-1}}\right) + 
\nonumber \\
& &v_{a_1}^{*}\left( \eta _1^{a_1}+A_{a_0}^{\;\;a_1}\eta
^{a_0}+Z_{\;\;a_2}^{a_1}\eta _2^{a_2}\right) +\nonumber \\
& &\sum\limits_{k=1}^bv_{a_{2k+1}}^{*}\left( \eta
_1^{a_{2k+1}}+A_{a_{2k}}^{\;\;a_{2k+1}}\eta
^{a_{2k}}+Z_{\;\;a_{2k+2}}^{a_{2k+1}}\eta _2^{a_{2k+2}}\right) +\cdots . 
\end{eqnarray}
The functions $\bar C_{\;\;b_{2k}c_{2i}}^{a_{2j}}$ and $\tilde
V_{b_{2k}}^{\;\;a_{2j}}$ are those involved with the irreducible gauge
algebra arising in the $L$-stage reducible case. In this way, we realized
the BRST quantization of arbitrary $L$-stage reducible first-class
Hamiltonian systems in an irreducible manner, i.e., without introducing
ghosts of ghosts. This completes our analysis.

\section{Example}

Here we exemplify the general theory exposed above in the case of abelian
one- and two-form gauge fields with Stueckelberg coupling. We start with the
Lagrangian action 
\begin{equation}
\label{70}S_0^L\left[ H^\mu ,A^{\mu \nu }\right] =-\int d^4x\left( \frac
1{12}F_{\mu \nu \rho }^2+\frac 14\left( MA_{\mu \nu }-F_{\mu \nu }\right)
^2\right) , 
\end{equation}
where $F_{\mu \nu }$ and $F_{\mu \nu \rho }$ denote the field strengths
associated with $H_\mu $, respectively, $A_{\mu \nu }$, and the notation $%
F_{\mu \nu \rho }^2$ signifies $F_{\mu \nu \rho }F^{\mu \nu \rho }$. (We
used a similar notation for the other square.) The system described by
action (\ref{70}) possesses the first-class constraints 
\begin{equation}
\label{71}G_i^{\left( 1\right) }\equiv \pi _{0i}\approx 0,\;G^{\left(
1\right) }\equiv \Pi _0\approx 0, 
\end{equation}
\begin{equation}
\label{72}G_i^{\left( 2\right) }\equiv -2\partial ^l\pi _{li}+M\Pi _i\approx
0,\;G^{\left( 2\right) }\equiv -\partial ^i\Pi _i\approx 0, 
\end{equation}
and the first-class Hamiltonian%
\begin{eqnarray}\label{73}
& &H=\int d^3x\left( -\pi _{ij}^2-
\frac{1}{2}\Pi _i^2+A^{0i}G_i^{\left( 2\right)
}+H^0G^{\left( 2\right) }+\right. \nonumber \\ 
& &\left. \frac 1{12}F_{ijk}^2+\frac 14\left( MA_{ij}-F_{ij}\right)
^2\right) . 
\end{eqnarray}
In (\ref{71}--\ref{73}), the $\pi $'s and $\Pi $'s are the canonical momenta
associated with the corresponding $A$'s and $H$'s. The first-class
constraints (\ref{72}) are first-stage reducible 
\begin{equation}
\label{74}\partial ^iG_i^{\left( 2\right) }+MG^{\left( 2\right) }=0, 
\end{equation}
with the reducibility functions 
\begin{equation}
\label{75}Z_{\;\;a_1}^{a_0}=\left( \partial ^i,M\right) . 
\end{equation}
The functions $A_{a_0}^{\;\;a_1}$ read 
\begin{equation}
\label{76}A_{a_0}^{\;\;a_1}=\left( 
\begin{array}{c}
-\partial _i \\ 
-M 
\end{array}
\right) , 
\end{equation}
such that $Z_{\;\;a_1}^{a_0}A_{a_0}^{\;\;b_1}=-\left( \partial ^i\partial
_i+M^2\right) $ is invertible. The variables $\left( y^{a_1},\pi
_{a_1}\right) $ will be denoted in this case by $\left( \varphi ,\pi \right) 
$. The irreducible first-class constraints are given by (\ref{71}) and 
\begin{equation}
\label{77}\gamma _i^{\left( 2\right) }\equiv -2\partial ^l\pi _{li}+M\Pi
_i-\partial _i\pi \approx 0,\;\gamma ^{\left( 2\right) }\equiv -\partial
^i\Pi _i-M\pi \approx 0, 
\end{equation}
while the first-class Hamiltonian $H^{\prime }$ (see (\ref{30})) reads 
\begin{equation}
\label{78}H^{\prime }=H+\int d^3x\left( -A^{0i}\partial _i\pi -MH^0\pi
\right) . 
\end{equation}
We introduce the pairs $\left( \varphi _1,\pi _1\right) $, $\left( \varphi
_2,\pi _2\right) $, and set the constraints 
\begin{equation}
\label{79}\gamma \equiv \pi -\pi _1\approx 0,\;\gamma ^{\prime }\equiv \pi
_2\approx 0. 
\end{equation}
The momentum $\pi $ is indeed a combination of the first-class constraints (%
\ref{77}) 
\begin{equation}
\label{80}\pi =-\frac 1{\left( \partial ^i\partial _i+M^2\right) }\left(
\partial ^i\gamma _i^{\left( 2\right) }+M\gamma ^{\left( 2\right) }\right) . 
\end{equation}
The first-class Hamiltonian with respect to (\ref{71}), (\ref{77}) and (\ref
{79}) has the form 
\begin{equation}
\label{81}H_0=H^{\prime }+\int d^3x\left( -\varphi _2\left( \partial
^i\gamma _i^{\left( 2\right) }+M\gamma ^{\left( 2\right) }\right) -\varphi
\pi _2\right) \equiv \int d^3xh_0. 
\end{equation}
The extended action%
\begin{eqnarray}\label{82}
& &S_0^E=\int d^4x\left( \dot A^{0i}\pi _{0i}+\dot A^{ij}\pi _{ij}+
\dot H^0\Pi
_0+\dot H^i\Pi _i+\dot \varphi \pi +\dot \varphi _1\pi _1+\right. 
\nonumber \\
& &\left. \dot \varphi _2\pi _2-h_0-u^iG_i^{\left( 1\right)
}-uG^{\left( 1\right) }-u^{\prime }\gamma -v^i\gamma _i^{\left( 2\right)
}-v\gamma ^{\left( 2\right) }-v^{\prime }\gamma ^{\prime }\right) , 
\end{eqnarray}
is invariant under the gauge transformations 
\begin{equation}
\label{83}\delta _\epsilon A^{0i}=\epsilon _1^i,\;\delta _\epsilon
H^0=\epsilon _1,\;\delta _\epsilon A^{ij}=\partial _{}^{\left[ i\right.
}\epsilon _2^{\left. j\right] },\;\delta _\epsilon H^i=\partial ^i\epsilon
_2+M\epsilon _2^i, 
\end{equation}
\begin{equation}
\label{84}\delta _\epsilon \varphi =\partial _i\epsilon _2^i-M\epsilon
_2+\tilde \epsilon _1,\;\delta _\epsilon \varphi _1=-\tilde \epsilon
_1,\;\delta _\epsilon \varphi _2=\tilde \epsilon _2,\;\delta _\epsilon
u=\dot \epsilon _1,\delta _\epsilon u^{\prime }=\stackrel{.}{\tilde \epsilon 
}_{1,} 
\end{equation}
\begin{equation}
\label{86}\delta _\epsilon v^i=\dot \epsilon _2^i-\partial ^i\tilde \epsilon
_2-\epsilon _1^i,\;\delta _\epsilon v=\dot \epsilon _2+M\tilde \epsilon
_2-\epsilon _1,\;\delta _\epsilon v^{\prime }=\stackrel{.}{\tilde \epsilon }%
_2+\partial _i\epsilon _2^i-M\epsilon _2+\tilde \epsilon _1, 
\end{equation}
the gauge variations of all the momenta being identically vanishing. In (\ref
{83}--\ref{86}) the gauge parameters $\epsilon _1^i$, $\epsilon _1$, $%
\epsilon _2^i$, $\epsilon _2$, $\tilde \epsilon _1$ and $\tilde \epsilon _2$
are respectively associated with the constraint functions $G_i^{\left(
1\right) }$, $G^{\left( 1\right) }$, $\gamma _i^{\left( 2\right) }$, $\gamma
^{\left( 2\right) }$, $\gamma $ and $\gamma ^{\prime }$. From (\ref{83}--\ref
{86}) we can derive the Lagrangian gauge transformations associated with the
irreducible theory (including, of course, the gauge transformations of the
original fields). In view of this we should consider a model of irreducible
Hamiltonian theory. In this light we assume that (\ref{71}) and the former
constraint in (\ref{79}) are primary, while (\ref{77}) and the latter
constraint from (\ref{79}) are secondary. Passing from the extended action (%
\ref{82}) to the corresponding total one (obtained by taking $v^i=0$, $v=0$
and $v^{\prime }=0$ in (\ref{82})) we derive its gauge invariances in the
standard manner. Indeed, the equations $v^i=0$, $v=0$ and $v^{\prime }=0$
imply $\delta _\epsilon v^i=0$, $\delta _\epsilon v=0$ and $\delta _\epsilon
v^{\prime }=0$. The last three equations lead via (\ref{86}) to 
\begin{equation}
\label{86a}\epsilon _1^i=\dot \epsilon _2^i-\partial ^i\tilde \epsilon
_2,\;\epsilon _1=\dot \epsilon _2+M\tilde \epsilon _2,\;\tilde \epsilon _1=-%
\stackrel{.}{\tilde \epsilon }_2-\partial _i\epsilon _2^i+M\epsilon _2. 
\end{equation}
Replacing $\epsilon _1^i$, $\epsilon _1$ and $\tilde \epsilon _1$ from (\ref
{86a}) in (\ref{83}--\ref{84}) we get 
\begin{equation}
\label{86b}\delta _\epsilon A^{0i}=\dot \epsilon _2^i-\partial ^i\tilde
\epsilon _2,\;\delta _\epsilon H^0=\dot \epsilon _2+M\tilde \epsilon
_2,\;\delta _\epsilon A^{ij}=\partial _{}^{\left[ i\right. }\epsilon
_2^{\left. j\right] },\;\delta _\epsilon H^i=\partial ^i\epsilon
_2+M\epsilon _2^i, 
\end{equation}
\begin{equation}
\label{86c}\delta _\epsilon \varphi =-\stackrel{.}{\tilde \epsilon }%
_2,\;\delta _\epsilon \varphi _1=\stackrel{.}{\tilde \epsilon }_2+\partial
_i\epsilon _2^i-M\epsilon _2,\;\delta _\epsilon \varphi _2=\tilde \epsilon
_2, 
\end{equation}
\begin{equation}
\label{86d}\delta _\epsilon u=\ddot \epsilon _2+M\stackrel{.}{\tilde
\epsilon }_2,\delta _\epsilon u^{\prime }=-\stackrel{..}{\tilde \epsilon }%
_2-\partial _i\dot \epsilon _2^i+M\dot \epsilon _2. 
\end{equation}
The Lagrangian action corresponding to the above total action coincides with
the original one, and its gauge transformations, which derive from (\ref{86b}%
--\ref{86c}), read as 
\begin{equation}
\label{86e}\delta _\epsilon A^{\mu \nu }=\partial ^\mu \epsilon ^\nu
-\partial ^\nu \epsilon ^\mu ,\;\delta _\epsilon H^\mu =\partial ^\mu
\epsilon +M\epsilon ^\mu ,\;\delta _\epsilon \varphi _1=\partial _\mu
\epsilon ^\mu -M\epsilon . 
\end{equation}
The gauge transformations for $\varphi $ and $\varphi _2$ were omitted as
these fields play the role of Lagrange multipliers (see (\ref{81})) and are
not relevant in the Lagrangian context. In order to write down (\ref{86e})
we employed the notations 
\begin{equation}
\label{86f}\epsilon ^\mu =\left( \tilde \epsilon _2,\epsilon _2^i\right)
,\;\epsilon =\epsilon _2. 
\end{equation}
In consequence, our formalism reproduces via (\ref{83}--\ref{86}) the
original gauge transformations and outputs some new gauge transformations
(for $\varphi _1$) that make the gauge transformation set (\ref{86e})
irreducible also at the Lagrangian level. The Lorentz covariance of the
gauge transformations (\ref{86e}) is due to the introduction in the theory
of the pairs $\left( \varphi _1,\pi _1\right) $ and $\left( \varphi _2,\pi
_2\right) $.

In the sequel we approach the antifield BRST treatment of (\ref{82}).
Straightforward calculation then yield the solution to the master equation 
\begin{eqnarray}\label{87}
& &S^E=S_0^E+\int d^4x\left( A_{0i}^{*}\eta _1^i+H_0^{*}\eta
_1+A_{ij}^{*}\partial _{}^{\left[ i\right. }\eta _2^{\left. j\right]
}+H_i^{*}\left( \partial ^i\eta _2+M\eta _2^i\right) +\right. \nonumber \\ 
& &\varphi ^{*}\left( \partial _i\eta _2^i-M\eta _2+\tilde \eta _1\right)
-\varphi _1^{*}\tilde \eta _1+\varphi _2^{*}\tilde \eta _2+u_i^{*}\dot \eta
_1^i+u^{*}\dot \eta _1+u^{\prime *}\stackrel{.}{\tilde \eta }_1+\nonumber \\ 
& &v_i^{*}\left( \dot \eta _2^i-\partial ^i\tilde \eta _2-\eta
_1^i\right) +v^{*}\left( \dot \eta _2+M\tilde \eta _2-\eta _1\right)
+\nonumber \\
& &\left. v^{\prime *}\left( \stackrel{.}{\tilde \eta }_2+
\partial _i\eta _2^i-M\eta _2+\tilde \eta _1\right) \right) . 
\end{eqnarray}
All the ghosts from (\ref{87}) have ghost number one, and all the antifields
ghost number minus one. We take the gauge fixing fermion 
\begin{eqnarray}\label{88}
& &\psi =\int d^4x\left( u_i^{*}\left( \partial _jA^{ji}+MH^i+\partial
^i\varphi _1\right) +u^{*}\left( \partial _iH^i-M\varphi _1\right) -\right. 
\nonumber \\
& &\left. u^{\prime *}\left( \partial _jA^{j0}+MH^0\right) \right) , 
\end{eqnarray}
which implements the irreducible gauge conditions $\partial
_jA^{ji}+MH^i+\partial ^i\varphi _1=0$, $\partial _iH^i-M\varphi _1=0$, and $%
\partial _jA^{j0}+MH^0=0$. After some computation we are led to the
gauge-fixed action 
\begin{eqnarray}\label{89}
& &S_\psi ^E=S_0^L+\int d^4x\left( B_\mu \left( \partial _\nu A^{\nu \mu
}+MH^\mu +\partial ^\mu \varphi _1\right) +b\left( \partial _\nu H^\nu
-M\varphi _1\right) +\right. \nonumber \\
& &\left. u_\mu ^{*}\left( \Box +M^2\right) \eta _2^\mu +u^{*}\left(
\Box +M^2\right) \eta _2\right) , 
\end{eqnarray}
such that the resulting path integral is given by 
\begin{equation}
\label{90}Z_\psi =\int {\cal D}A^{\nu \mu }{\cal D}H^\mu {\cal D}B_\mu {\cal %
D}\varphi _1{\cal D}u_\mu ^{*}{\cal D}\eta _2^\mu {\cal D}u^{*}{\cal D}\eta
_2\exp iS_\psi ^E. 
\end{equation}
In (\ref{89}-\ref{90}) we employed the identifications 
\begin{equation}
\label{91}B_\mu =\left( \pi _1,\pi _{0i}\right) ,\;b=\Pi _{0,}\;u_\mu
^{*}=\left( -u^{\prime *},u_j^{*}\right) \;\eta _2^\mu =\left( \tilde \eta
_2,\eta _2^j\right) . 
\end{equation}
One can check that there are no residual gauge invariances in action (\ref
{89}). Moreover, the gauge-fixed action (\ref{89}) is Lorentz covariant.
This is due precisely to the introduction in the theory of the pairs $\left(
\varphi _1,\pi _1\right) $ and $\left( \varphi _2,\pi _2\right) $ subject to
the constraints (\ref{79}).

While the gauge-fixing fermion (\ref{88}) is useful in getting the covariant
path integral (\ref{90}), the fermion 
\begin{eqnarray}\label{92}
& &\psi ^{\prime }=\int d^4x\left( u_i^{*}\left( \partial _jA^{ji}-\dot
A^{0i}+\partial ^i\varphi _1\right) +\right. \nonumber \\  
& &\left. u^{*}\left( \partial _iH^i-\dot H^0\right) -u^{\prime
*}\left( \partial _jA^{j0}-\dot \varphi _1\right) \right) ,
\end{eqnarray}
is appropriate in order to make the reduction to the physical degrees of
freedom in the path integral. Starting with the solution (\ref{87}) and on
behalf of (\ref{92}) we find after some computation the path integral over
physical degrees of freedom for the model under consideration of the form%
\begin{eqnarray}\label{93}
& &Z_{\psi ^{\prime }}=
\int {\cal D}A^{ij}{\cal D}\pi _{ij}{\cal D}H^i{\cal D}%
\Pi _i{\cal D}\varphi _1{\cal D}\pi _1\delta \left( \partial ^j\pi
_{ji}+\partial _i\pi _1\right) \times \nonumber \\  
& &\delta \left( \partial _jA^{ji}+\partial ^i\varphi _1\right)
\delta \left( \partial ^i\Pi _i\right) \delta \left( \partial _iH^i\right)
\exp i\bar S_{\psi ^{\prime }},
\end{eqnarray}
where $\bar S_{\psi ^{\prime }}$ is given by 
\begin{equation}
\label{94}\bar S_{\psi ^{\prime }}=\int d^4x\left( \dot A^{ij}\pi _{ij}+\dot
H^i\Pi _i+\pi _{ij}^2+\frac 12\Pi _i^2-\frac 1{12}F_{ijk}^2-\frac 14\left(
MA_{ij}-F_{ij}\right) ^2\right) . 
\end{equation}
The delta functions from the constraint functions and their gauge conditions
in the path integral (\ref{93}) show that the independent fields and momenta
are precisely the transverse components of $H^i$ and $\Pi _i$ and also the
longitudinal components of $A^{ij}$ and $\pi _{ij}$. It is clear that the
conditions $\partial ^i\Pi _i=0$ and $\partial _iH^i=0$ restrict the
integration only over the two transverse degrees of freedom for the vector
fields and their momenta (typically for electromagnetism). Related to the
remaining conditions from the measure of (\ref{93}), it can be shown that
they enforce the longitudinal parts as independent components of the tensor
fields and their momenta. Indeed, $A^{ij}$ and $\pi _{ij}$ can be decomposed
into longitudinal and transverse components 
\begin{equation}
\label{95}A_{ij}=\partial _iA_j^T-\partial _jA_i^T+\varepsilon
_{ijk}\partial ^kA^L,\;\pi _{ij}=\partial _i\pi _j^T-\partial _j\pi
_i^T+\varepsilon _{ijk}\partial ^k\pi ^L, 
\end{equation}
where the transverse components satisfy $\partial ^iA_i^T=0$ and $\partial
^i\pi _i^T=0$. Then, the conditions $\partial _jA^{ji}+\partial ^i\varphi
_1=0$ and $\partial ^j\pi _{ji}+\partial _i\pi _1=0$ imply via (\ref{95})
that 
\begin{equation}
\label{96}\partial ^i\partial _iA_j^T+\partial _j\varphi _1=0,\;\partial
^i\partial _i\pi _j^T+\partial _j\pi _1=0, 
\end{equation}
hence 
\begin{equation}
\label{97}A_j^T=-\frac{\partial _j}{\triangle }\varphi _1,\;\pi _j^T=-\frac{%
\partial _j}{\triangle }\pi _1. 
\end{equation}
On the other hand, from (\ref{96}) it follows that $\partial ^i\partial
_i\varphi _1=0$ and $\partial ^i\partial _i\pi _1=0$, which then yield $%
\varphi _1=0$, $\pi _1=0$ by virtue of the boundary conditions for the
unphysical degrees of freedom $\left( \varphi _1,\pi _1\right) $ (vacuum to
vacuum). Inserting the last relations back in (\ref{97}) we find that the
conditions checked by the tensor fields and their momenta lead to $A_j^T=0$
and $\pi _j^T=0$, so the only physical degrees of freedom are described by
the longitudinal pair $\left( A^L,\pi ^L\right) $. In this way the
conditions implemented in the measure of (\ref{93}) lead to transverse
degrees of freedom for the vector fields, respectively to a longitudinal one
for the tensor fields, like in the reducible approach. This completes the
analysis of the investigated model.

\section{Conclusion}

In conclusion, we succeeded in giving a systematic irreducible procedure of
quantizing reducible first-class Hamiltonian systems accordingly the
antifield BRST method. This new result was inferred by means of constructing
an irreducible first-class Hamiltonian theory in a larger phase-space that
remains physically equivalent to the original redundant one. The above
equivalence makes legitimate the replacement of the quantization of the
reducible theory by that of the irreducible system. As a consequence of our
irreducible approach, the ghosts of ghosts, their antifields, as well as the
pyramidal structure of auxiliary fields are no longer necessary. We further
illustrate in detail the theoretical part of the paper in the case of the
Stueckelberg coupled abelian one- and two-form gauge fields.

\end{document}